\documentclass[11pt]{article}
\usepackage{zeller-moriond,epsfig}

\def\np#1#2#3   {{ Nucl. Phys.} {\bf#1}, #2 (#3) }
\def\pcps#1#2#3 {{ Proc. Cam. Phil. Soc.} {\bf#1}, #2 (#3) }
\def\pl#1#2#3   {{ Phys. Lett.} {\bf#1}, #2 (#3) }
\def\plc#1#2#3   {{ Phys. Lett.} {\bf#1}, #2 (#3) }
\def\prep#1#2#3 {{ Phys. Rep.} {\bf#1}, #2 (#3) }
\def\prev#1#2#3 {{ Phys. Rev.} {\bf#1}, #2 (#3) }
\def\prl#1#2#3  {{ Phys. Rev. Lett.} {\bf#1}, #2 (#3) }
\def\prs#1#2#3  {{ Proc. Roy. Soc.} {\bf#1}, #2 (#3) }
\def\ptp#1#2#3  {{ Prog. Th. Phys.} {\bf#1}, #2 (#3) }
\def\rmp#1#2#3  {{ Rev. Mod. Phys.} {\bf#1}, #2 (#3) }
\def\rpp#1#2#3  {{ Rep. Prog. Phys.} {\bf#1}, #2 (#3) }
\def\zp#1#2#3   {{ Zeit. Phys.} {\bf#1}, #2 (#3) }
\def\epj#1#2#3   {{ Eur. Phys. Jour.} {\bf#1}, #2 (#3) }
\def\nim#1#2#3   {{ Nucl. Instr. Meth.} {\bf#1}, #2 (#3) }

\newcommand{\rmt}{\rm\textstyle}
\newcommand{\rms}{\rm\scriptstyle}
\newcommand{\stw}{\mbox{$\sin^2\theta_W$}}     
\newcommand{\stww}{\mbox{$\sin^4\theta_W$}}     
\newcommand{\mw}{\mbox{$M_W$}}     
\newcommand{\mtop}{\mbox{$m_t$}}
\newcommand{\mhiggs}{\mbox{$m_H$}}
\newcommand{\nub}{{\overline{\nu}}}          
\newcommand{\Rnu}{\mbox{$R^{\nu}$}}
\newcommand{\Rnub}{\mbox{$R^{\nub}$}}
\newcommand{\Rmeasnu}{\mbox{$R_{\rms exp}^{\nu}$}}
\newcommand{\Rmeasnub}{\mbox{$R_{\rms exp}^{\nub}$}}
\newcommand{\gLeff}{\mbox{$g_L^{\rms eff}$}}
\newcommand{\gReff}{\mbox{$g_R^{\rms eff}$}}
\newcommand{\ubar}{\overline{u}}     
\newcommand{\dbar}{\overline{d}}     
\newcommand{\sbar}{\overline{s}}     
\newcommand{\cbar}{\overline{c}}     
\newcommand{\pbar}{\overline{p}}  
\newcommand{\numu}{\nu_\mu}
\newcommand{\numubar}{\overline{\nu}_\mu} 

\begin{document}
\vspace*{4cm}
\title{A DEPARTURE FROM PREDICTION: ELECTROWEAK PHYSICS AT NUTEV}

\author{G.~P.~Zeller \\ (for the NuTeV Collaboration) }

\address{Northwestern University, Evanston, IL, 60208, USA}


\maketitle\abstracts{The NuTeV collaboration has extracted the electroweak 
parameter, $\stw$, from the measurement of the ratios of neutral current
to charged current neutrino and antineutrino deep inelastic scattering
interactions. We find that our measurement, while in agreement with previous 
neutrino electroweak measurements, is not consistent with the prediction from 
global electroweak fits. To facilitate interpretation of the result, a model 
independent analysis is presented and possible explanations are discussed.}

\section*{Introduction}

In deep inelastic neutrino-nucleon scattering, the weak mixing angle can
be extracted from the ratio of neutral current (NC) to charged current (CC)
total cross sections~\cite{llewellyn-smith}:
 
\begin{eqnarray*}
R^{\nu} \equiv \frac{\sigma(\nu_{\mu}N\rightarrow\nu_{\mu}X)}
                 {\sigma(\nu_{\mu}N\rightarrow\mu^-X)}
        &=& \frac{\sigma_{NC}^\nu}{\sigma_{CC}^\nu}
         = g_L^2 + r\,g_R^2 \\
        &=& \frac{1}{2}-\sin^2\theta_W+\frac{5}{9}\,(1+r)\,\sin^4\theta_W \\
R^{\nub} \equiv \frac{\sigma(\nub_{\mu}N\rightarrow\nub_{\mu}X)}
                 {\sigma(\nub_{\mu}N\rightarrow\mu^+X)}
        &=& \frac{\sigma_{NC}^{\nub}}{\sigma_{CC}^{\nub}}
         = g_L^2 + \frac{1}{r}\, g_R^2 \\
        &=& \frac{1}{2}-\sin^2\theta_W+\frac{5}{9}\,\left(1+ \frac{1}{r}\right)
           \,\sin^4\theta_W
\end{eqnarray*}

\noindent
where $r=\sigma_{CC}^{\nub}/\sigma_{CC}^\nu$ and $g_L^2=1/2-\stw+5/9\,\stww$
and $g_R^2=5/9\,\stww$ are the left and right handed isoscalar quark 
couplings, respectively. The above relations are, of course, exact only for 
tree level scattering off an isoscalar target composed of light quarks. 
Necessary adjustments to this idealized model include corrections for the 
non--isoscalar target, quark mixing, radiative effects, higher--twist 
processes, the longitudinal structure function ($R_L$), the W and Z 
propagators, and the heavy quark content of the nucleon (charm and strange). 
Unfortunately, previous determinations of $\stw$ measured using $\Rnu$ 
suffered from large theoretical uncertainties associated with heavy quark 
production thresholds mainly affecting the CC denominator. These 
uncertainties, resulting from imprecise knowledge of the charm quark mass, 
dominated the CCFR measurement~\cite{ccfr} and ultimately limited the 
precision of neutrino measurements of electroweak parameters. For example, 
combining the five most precise neutrino--nucleon measurements yielded a 
value of $\stw^{\nu N} \equiv 1 - M_W^2/M_Z^2 = 0.2277 \pm 
0.0036$,~\cite{ichep} thereby implying an equivalent W mass error of 190 
MeV. 

The Paschos--Wolfenstein combination~\cite{paschos} provides an alternative 
method for determining $\stw$ that is much less dependent on the details of 
charm production and other sources of model uncertainty:

\vspace{-0.25in}
\begin{eqnarray*}
R^{-}
= \frac{\sigma^{\nu}_{NC}-\sigma^{\bar \nu}_{NC}}
       {\sigma^{\nu}_{CC}-\sigma^{\bar \nu}_{CC}}
= \frac{R^{\nu}-rR^{\nub}}{1-r}=\frac{1}{2}-\sin^2\theta_W
\end{eqnarray*}

\noindent
Under the assumption that the neutrino--quark and antineutrino--antiquark
cross sections are equal, use of the Paschos--Wolfenstein relation removes 
the effects of sea quark scattering which dominate the low $x$ cross section.
As a result, $R^-$ is much less sensitive to heavy quark processes provided 
these contributions are the same for neutrinos and antineutrinos. The only 
remaining charm--producing contributors are $d_v$ quarks which are not only 
Cabibbo suppressed but are also at higher fractional momentum, $x$, where 
the mass suppression is less of an effect.

Inspired by the Paschos--Wolfenstein technique, the measurement presented here
extracts electroweak parameters from neutrino and antineutrino deep inelastic
scattering reactions. However, NuTeV does not measure cross section ratios, 
such as those appearing in the above expressions ($R^-$, $\Rnu$, $\Rnub$)
because of our inability to measure NC interactions down 
to zero recoil energy and because of the presence of experimental cuts, 
backgrounds, and detector acceptance. NuTeV instead measures experimental 
ratios of short to long events, $\Rmeasnu$ and $\Rmeasnub$. A detailed Monte 
Carlo simulation of the experiment then predicts these ratios and their 
dependence on electroweak parameters~\cite{mythesis}. In the end, the NuTeV 
measurement has comparable precision to other experimental tests. Because 
neutrino scattering is a different physical process, NuTeV is 
sensitive to different new physics. In addition, NuTeV provides a precise 
measurement of NC neutrino couplings (the only other precise measurement is 
from the LEP I invisible line width), a measurement of processes at moderate 
space--like momentum transfers (as opposed to large time--like transfers 
probed at collider experiments), as well as a precise determination of the 
parameters of the model itself ($\stw$, $\mw$, $\rho_0$, $g_L^2$, and $g_R^2$).

\section*{Results}

From the high statistics samples of separately collected neutrino and
antineutrino events and assuming the standard model, NuTeV finds:
 
\vspace{-0.05in}
\begin{eqnarray*}
 \stw^{\nu N} \equiv 1 - M_W^2/M_Z^2 
      &=& 0.2277\pm0.0013 \: ({\rmt stat})\pm0.0009\:({\rmt syst}) \\
      &-&0.00022\times\left(\frac{\mtop^2-(175 \: \mathrm{GeV})^2}{(50 \: \mathrm{GeV})^2}\right) \\
      &+&0.00032\times \ln\left(\frac{\mhiggs}{150 \: \mathrm{GeV}}\right)
\end{eqnarray*}
 
\noindent
with the small residual dependence on $\mtop$ and $\mhiggs$ 
resulting from leading terms in the one--loop electroweak radiative 
corrections to the W and Z self energies~\cite{radcorr}.
The result lies three standard deviations above the prediction from 
the global electroweak fit, $0.2227 \pm 0.0004$~\cite{LEPEWWG,Martin}. 
The measurement is currently the most precise determination of $\stw$ 
in neutrino--nucleon scattering, surpassing its predecessors by a factor 
of two in precision, and is statistics--dominated. Within the standard 
model, the NuTeV measurement of $\stw$ indirectly determines the W boson 
mass, $M_W = 80.14 \pm 0.08$ GeV, with a precision comparable to
individual direct measurements from high energy $e^+e^-$ and $p\pbar$ 
colliders; however, the nearly $3.5\sigma$ deviation from the directly 
measured W mass, $\mw=80.45\pm0.04$ GeV, makes it especially difficult 
to explain the NuTeV result in terms of oblique radiative 
corrections~\cite{rosner}.

Relaxing the standard model assumptions, a model independent analysis 
recasts the same data into a measurement of effective left and right 
handed neutral current quark couplings. NuTeV measures:

\vspace{-0.3in}
\begin{eqnarray*}
   (\gLeff)^2 &=& 0.3001 \pm 0.0014 \\
   (\gReff)^2 &=& 0.0308 \pm 0.0011
\end{eqnarray*}
 
\noindent
with a correlation coefficient of $-0.017$. Comparing these couplings to 
their standard model values~\cite{LEPEWWG}, 
$(\gLeff)_{\mathrm{SM}}^2=0.3042$ and $(\gReff)_{\mathrm{SM}}^2=0.0301$,
indicates that while the right handed coupling appears to be compatible
with the prediction, the NuTeV data clearly prefer a smaller left handed 
effective coupling. 



Lying $3\sigma$ above the prediction of the standard electroweak theory,
the NuTeV $\stw$ result is surprising, however it is not immediately
apparent what the cause of the discrepancy might be. In the following 
sections, we discuss the impact of the NuTeV result on global standard 
model fits, the plausibility of various explanations, and the prospects for 
future measurements of $\stw$ at low energy.

\section*{Impact on Standard Model Fit}

\begin{figure}[ht]	
\mbox{
\begin{minipage}{0.45\textwidth}
\centerline{\mbox{\psfig{figure=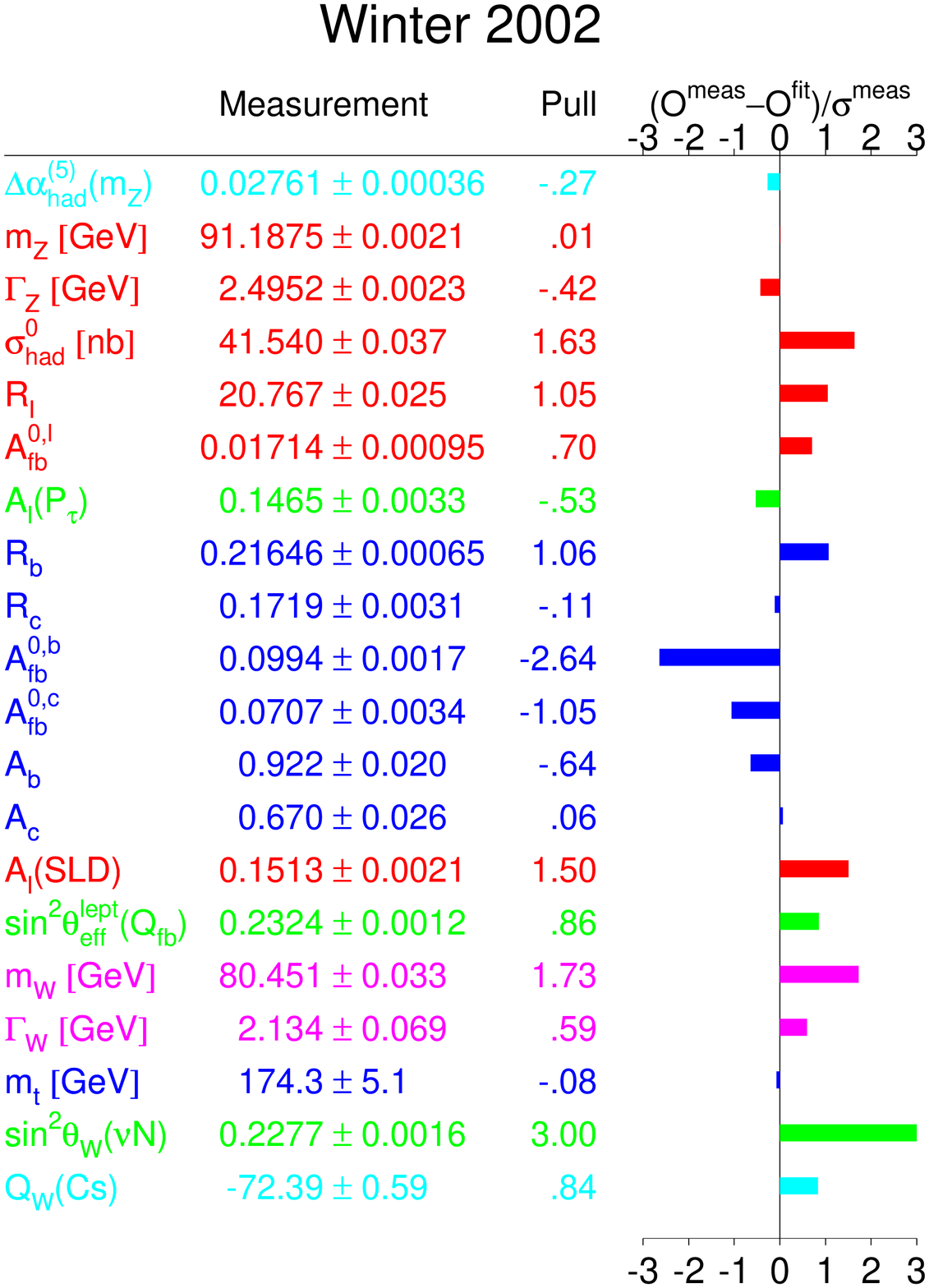,width=7.cm}}}
\vspace{-0.3in}
\caption{The current global electroweak fit including the NuTeV $\stw$ result. 
         The horizontal bars indicate the pull of each measurement, in 
         standard deviations, from its standard model expectation. Plot is 
         courtesy of the LEPEWWG.}
\label{fig:lepewwg-pull}
\end{minipage}\hspace*{0.4in}	
\begin{minipage}{0.45\textwidth}
\centerline{\mbox{\psfig{figure=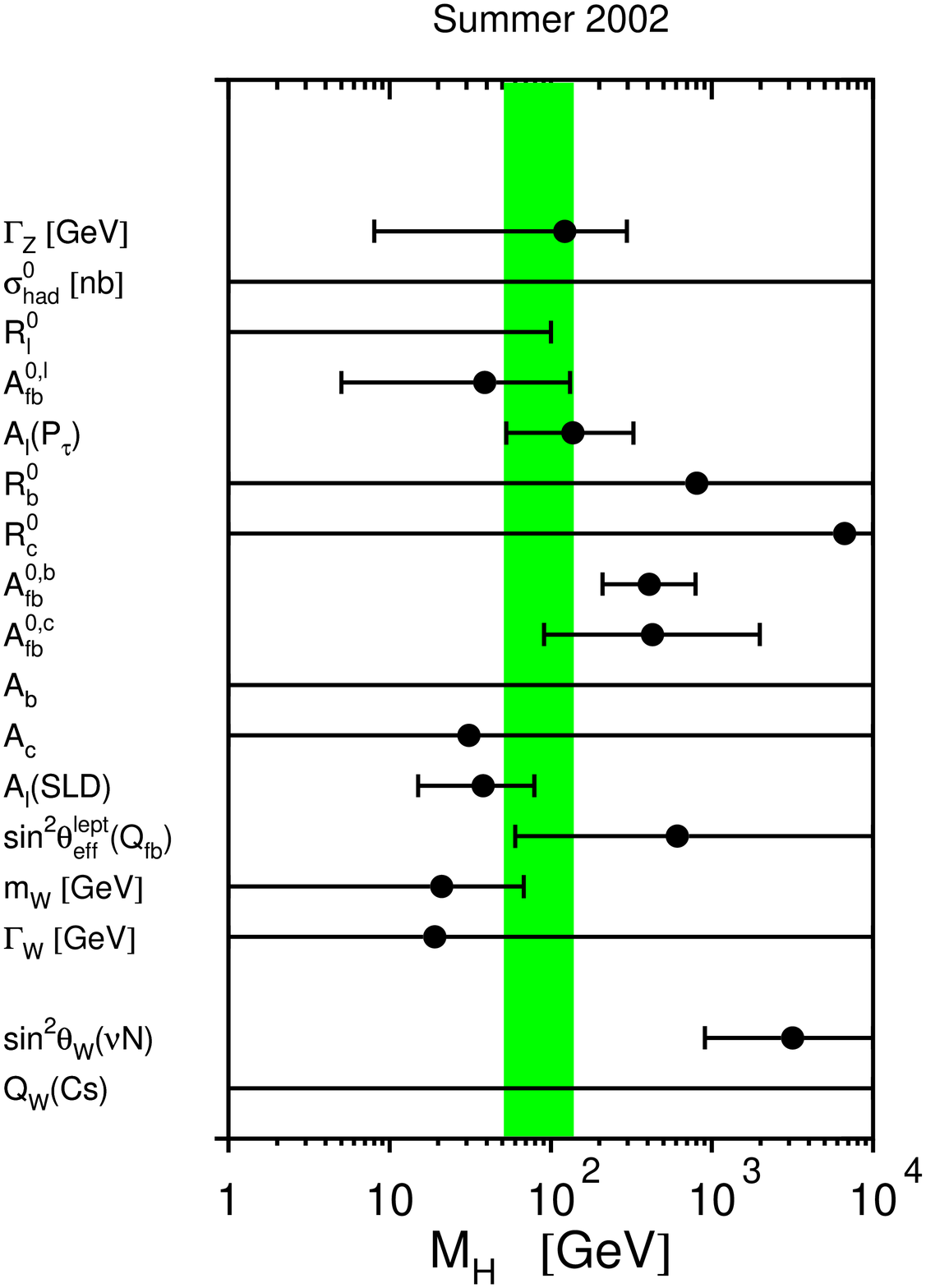,width=7.cm}}}
\vspace{-0.3in}
 \caption{Sensitivity of the precision electroweak data to $\mhiggs$.
          Most of the data is consistent with a low $\mhiggs$, except
          for $A_{FB}^{0,b}$ and NuTeV $\stw$. Plot is courtesy of the 
          LEPEWWG.}
 \label{fig:lepewwg-higgs}
\end{minipage}
}
\end{figure}

Figure~\ref{fig:lepewwg-pull} exhibits the results of the LEP Electroweak 
Working Group (LEPEWWG) global fit~\cite{LEPEWWG} to all precision electroweak 
data including the NuTeV measurement of $\stw$. The largest pulls are coming 
from the NuTeV $\stw$ result and the LEP II measurement of $A_{FB}^{0,b}$,
both of which favor a large Higgs mass (Figure~\ref{fig:lepewwg-higgs}).
The inclusion of NuTeV in the standard model fit increases the global 
$\chi^2$/dof to 28.8/15. The probability of the $\chi^2$ being worse than 
28.8 is only 1.7$\%$. If one arbitrarily excludes the NuTeV results, the fit 
improves to a probability of 14.3$\%$ ($\chi^2$/dof = 19.6/14), which itself
is marginalized by the $3\sigma$ discrepancy between the two most precise 
determinations of $\stw$ at the Z pole: the leptonic measurement, $A_{LR}$ 
at SLD, and the hadronic measurement, $A_{FB}^{0,b}$ at LEP.

These results should, of course, be interpreted with caution. Discarding one 
or two measurements can improve the fit, but at the same time drastically 
change the predicted Higgs boson mass. If the two most discrepant 
measurements, $A_{FB}^{0,b}$ and NuTeV $\stw$, are arbitrarily removed from 
the fit, the global $\chi^2$/dof improves to 6.84/9, a robust 65$\%$ 
probability~\cite{chanowitz-private}; however, the favored value of the Higgs 
mass drops to 43 GeV~\cite{chanowitz}, well below the direct search limits 
set by the non--discovery of the Higgs at LEP II, $\mhiggs>114$ GeV.

Motivated by the large standard model fit $\chi^2$, we explore possible 
explanations for the NuTeV results in the following sections. In particular, 
we consider the effects of nuclear shadowing, isospin violating parton 
distribution functions, asymmetries in the nucleon strange sea, and 
additional Z$^\prime$ bosons.

\section*{Nuclear Shadowing}

If nuclear shadowing were significantly different for NC and CC neutrino 
interactions, such an effect would impact NuTeV's measurement of $\stw$.
In a recent comment~\cite{thomas-comment}, Miller and Thomas consider
a particular vector meson dominance (VMD) shadowing model that they claim
is capable of accounting for the entire NuTeV discrepancy. However, as 
shadowing within the VMD model is weaker for $Z^0$ exchange than for 
$W^{\pm}$ exchange, 
the predictions for $\Rnu$ and $\Rnub$ are thereby increased for a portion of 
the NuTeV data in the low $Q^2$ shadowing region. The effect has the 
{\em wrong sign}, since NuTeV measures ratios for neutrino and antineutrino 
scattering processes, $\Rmeasnu$ and $\Rmeasnub$, which are both smaller 
than expected. More generally, because any model of differing neutral and 
charged current nuclear shadowing will change $\Rmeasnu$ and $\Rmeasnub$
more than $R^-$, it is unlikely that any such model could explain the
discrepancy in NuTeV's measurement of $\stw$. 

\section*{Isospin Violations}

The NuTeV result is extracted assuming isospin symmetry in the nucleon, 
$u^p=d^n$, $d^p=u^n$, $\ubar^p=\dbar^n$, and $\dbar^p=\ubar^n$. 
While all global parton distribution fits (CTEQ, GRV, MRST) are performed 
under this assumption, the NuTeV analysis is sensitive because of the need 
to assign $u$ and $d$ flavors (which possess different NC couplings) to the 
neutrino scatterers. Several classes of non--perturbative models have 
calculated the potential effect of isospin violation in the 
nucleon~\cite{sather,thomas,cao}. Estimating the effect of the single quark 
mass difference ($m_d-m_u = 4.3$ MeV), the earliest calculation~\cite{sather} 
predicts a large $-0.0020$ shift 
in $\stw^{\mathrm{NuTeV}}$, which could account for roughly $40\%$ 
of the observed discrepancy. However, more complete calculations that include 
differences in the nucleon masses ($m_n-m_p=1.3$ MeV), diquark masses 
($m_{dd}-m_{uu}$), and nucleon radii predict much smaller shifts in the 
result. For example, the Thomas {\em et al.} bag model 
calculation~\cite{thomas} predicts $\delta\,\stw^{\mathrm{NuTeV}}
=-0.0001$ as a result of the cancellation of opposing shifts at low and 
high $x$. A meson cloud model prediction~\cite{cao} yields a similarly small 
$+0.0002$ shift in the NuTeV measurement. To shift the NuTeV $\stw$ value down
to its standard model expectation would require isospin violation at the
level of $\int x\,d^p_v(x) - x\,u^n_v(x) \: dx \sim 0.01$ (or $5\%$ of
$\int x\,d^p_v(x) + x\,u^n_v(x) \: dx$)~\cite{nc-asym}. While the more recent 
calculations do not suggest large isospin violation, such a possibility cannot
be firmly excluded as a potential explanation for the NuTeV results. However, 
a nucleon isospin violating model which successfully accounts for the NuTeV 
discrepancy needs to be evaluated in the context of a global fit so as not 
to violate existing experimental data in the attempt to accommodate NuTeV.

\section*{Strange Sea Asymmetry}

The NuTeV analysis also assumes that the strange and anti--strange seas are 
symmetric, $s(x) = \sbar(x)$; however it has been noted that non--perturbative
QCD processes can potentially generate a momentum asymmetry between the 
strange and anti--strange seas~\cite{ssbar-theory}. Such an asymmetry can be 
directly measured using the same parton distribution formalism and cross 
section model as were employed in the $\stw$ measurement. Recall
that in neutrino scattering, dimuon events are a clean signature of charged 
current charm production ($\numu \, s \rightarrow \mu^- c$ and 
$\numubar \sbar \rightarrow \mu^+ \cbar$) and hence allow independent 
extractions of strange and anti--strange quark distributions. Leading order 
fits to the NuTeV neutrino and antineutrino dimuon data samples~\cite{max} 
yield a negative momentum asymmetry:

\begin{equation}
   \int x\,s(x) - x\,\sbar(x) \: dx = -0.0027 \pm 0.0013
\end{equation}

\noindent
and a corresponding {\em increase} in the NuTeV measurement of $\stw$:

\begin{equation}
   \stw = 0.2297 \pm 0.0019 
\end{equation}

\noindent
when compared to the result extracted assuming $s(x)=\sbar(x)$, 
$\stw=0.2277 \pm 0.0016$. Including the measured strange sea asymmetry
{\em increases} the NuTeV discrepancy with the standard model to $3.7\sigma$ 
significance, and hence, this is not a likely explanation. To explain the 
NuTeV $\stw$ result would require a strange sea asymmetry,
$\int x\,s(x) - x\,\sbar(x) \: dx \sim +0.007$, that is roughly $30\%$ of 
$\int x\,s(x) + x\,\sbar(x) \: dx$ and is in the opposite 
direction~\cite{nc-asym}.

\section*{Extra Z$^\prime$ Bosons}

In addition to evaluating the effects of unexpected parton 
asymmetries~\cite{nc-asym}, we also consider several non--standard physics 
cases. The existence of an additional Z boson would impact the NuTeV 
measurement by shifting the effective neutrino--quark couplings away from 
their standard model values. These shifts can arise from 
both pure Z$^\prime$ exchange as well as from Z--Z$^\prime$ mixing. A popular
class of Z$^\prime$ models involves the introduction of extra U(1) symmetries.
The E$_6$ model in particular has been considered as a candidate for grand 
unified theories. In this specific model, the coupling shifts are well 
determined~\cite{zprime}, however because the NuTeV result requires an 
enhancement in the effective left--handed quark couplings, it is difficult to 
explain the entire discrepancy with the inclusion of such a Z$^\prime$. While 
this specific model can produce large right--handed coupling shifts, 
appreciable Z--Z$^\prime$ mixing is required to induce sizable shifts in the 
left--handed couplings. The size of the mixing is severely limited, at the 
$\sim10^{-3}$ level, by measurements from LEP and SLD~\cite{zmixing}, hence 
making it difficult to accommodate the NuTeV measurement. On the other hand, 
it is possible to explain the entire NuTeV discrepancy with the inclusion of 
an ``almost'' sequential~\footnote{A Z$^\prime$ with standard couplings but 
which interferes destructively with the standard model Z.} Z$^\prime$ with 
a mass in the $1.2^{+0.3}_{-0.2}$ TeV range. The present limits from 
Run I CDF and D\O\ direct searches set 
$M_{Z^\prime_{\mathrm{SM}}}\stackrel{>}{\sim} 700$ GeV at $95\%$ confidence 
level~\cite{tevzprime}. Both the Tevatron Run II and the LHC offer the hope of 
discovering a Z$^\prime$ boson should it exist. Several authors have also 
recently discussed the NuTeV results in the context of other $U(1)$ extensions
and have found TeV scale Z$^\prime$'s with specific couplings capable of 
explaining the NuTeV discrepancy~\cite{extraU1,extraU2}.

\section*{Anomalous Neutrino NC Interaction}

Finally, while such a solution is not model--independent or unique, 
it is interesting to interpret the entire NuTeV discrepancy as a 
deviation in the overall NC coupling strength $\rho_0$. The result is 
a neutral current rate that is $1\%$ lower than the standard model 
expectation at almost $3\sigma$ significance:

\vspace{-0.2in}
\begin{equation}
  \rho^2_0 = 0.9884 \pm 0.0026\,(\mathrm{stat}) \pm 0.0032\,(\mathrm{syst})
\end{equation}

\noindent
Unlike in the NuTeV fit for $\stw$, both the neutrino and antineutrino data are
sensitive to $\rho_0$, so there is less control over the charm production 
uncertainties, and the systematics are therefore much larger.

\begin{figure}[h]
\centerline{\mbox{\psfig{figure=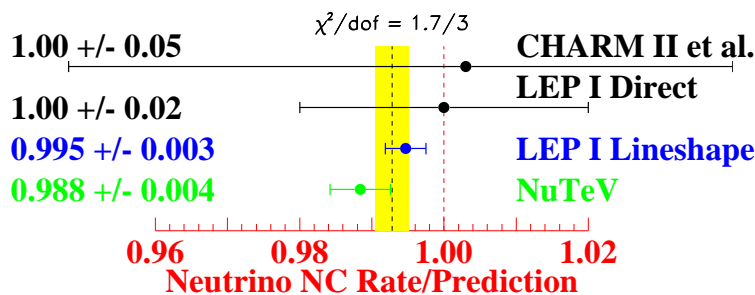,width=12.cm}}}
 \vspace{-0.2in}
 \caption{Experimental constraints on neutrino neutral current interaction
          rates relative to the standard model expectation. The two precise 
          measurements, LEP I $\Gamma(Z \rightarrow \nu \nub)$ and NuTeV 
          $\rho_0^2$, are both below expectation.}
 \label{fig:nc-rate}
\end{figure}

Figure~\ref{fig:nc-rate} displays the NuTeV result in comparison to all 
existing neutrino measurements. The only other precise experimental constraint
is the LEP I measurement of Z decays into invisible channels from which the 
number of light neutrino species can be deduced. The LEP I result, 
$N_{\nu} = 3 \cdot 
\frac{\Gamma_{\mathrm{meas}}(Z\rightarrow \nu\nub)}{\Gamma_{\mathrm{SM}}
(Z\rightarrow \nu\nub)} = 3 \cdot (0.9947 \pm 0.0028)$, is two standard
deviations shy of the three known neutrino species \cite{LEPEWWG}. Given this 
particular interpretation, one might suspect the neutral current couplings of 
neutrinos since the only two precise measurements are both lower than the 
standard model expectation. In fact, models capable of accommodating both 
the LEP I neutrino deficit and the NuTeV result have been recently proposed 
in the literature~\cite{ncrate}.

\subsection*{The Low Energy Future}

NuTeV was dismantled several years after data--taking and holds no hope of 
remeasuring electroweak parameters in neutrino scattering. While atomic parity
violation measurements~\cite{apv} will hopefully continue to improve,
in addition, two future experiments are preparing to also test the low energy 
prediction of $\stw$. An $e^+e^-$ M$\o$ller scattering experiment, E158 at 
SLAC~\cite{e158}, and a polarized electron--proton scattering experiment, 
QWEAK at Jefferson Lab~\cite{qweak}, both plan to probe this low Q$^2$ regime 
in the near future. If they too observe a significant deviation from the 
predicted $\stw$ scaling, this would provide striking evidence for new 
physics. However, if the deviation in the NuTeV measurement somehow resulted 
from new physics specific only to the neutrino or muon sector (i.e. that is 
not flavor universal), then the discrepancy would not manifest itself in 
these two future experiments.

\section*{Conclusions}

NuTeV has achieved the precision to be an important test of the electroweak
standard model. By measuring ratios of neutral to charged current interactions,
NuTeV has precisely determined $\stw$ and has found a discrepancy 
of three standard deviations from the standard model expectation. Models 
for new physics that are capable of explaining the NuTeV results tend to 
be exotic, but hopefully either future low or high energy experiments will 
provide a clue to the source of the discrepancy.

\section*{Acknowledgments}

We are thankful for the assistance of the staff of the Fermilab Beams, 
Computing, and Particle Physics Divisions as well as the support of the
U.S. Department of Energy and the Alfred P. Sloan Foundation. In addition,
we thank Stan Brodsky, Michael Chanowitz, Sacha Davidson, Jens Erler, 
Paolo Gambino, Martin Grunewald, Paul Langacker, Michael Peskin, Jon Rosner, 
and Tony Thomas for their useful input and numerous discussions.


\end{document}